\documentclass[aps,tightenlines,showpacs,amsmath,amssymb]{revtex4}

\usepackage[dvips]{graphicx}
\usepackage{dcolumn}
\usepackage{bm}

\begin{document}

\preprint{}

\title{Fluctuation Induced Instabilities in Front Propagation up a Co-Moving Reaction Gradient in Two Dimensions}

\author{C. Scott Wylie}
\email{cwylie@physics.ucsd.edu}
\affiliation{
Center for Theoretical Biological Physics, University of California, San Diego, 9500 Gilman Drive, La Jolla, CA 92093-0319
}
\author{Herbert Levine}
\email{hlevine@ucsd.edu}
\affiliation{
Center for Theoretical Biological Physics, University of California, San Diego, 9500 Gilman Drive, La Jolla, CA 92093-0319
}
\author{David A. Kessler}
\email{kessler@dave.ph.biu.ac.il}
 \affiliation{
Dept. of Physics, Bar-Ilan University, Ramat-Gan, IL52900 Israel}

\date{\today}

\begin{abstract}
We study 2D fronts propagating up a co-moving reaction rate gradient in finite number reaction-diffusion systems.  We show that in a 2D rectangular channel, planar solutions to the deterministic mean-field equation are stable with respect to deviations from planarity.  We argue that planar fronts in the corresponding stochastic system, on the other hand, are unstable if the channel width exceeds a critical value.  Furthermore, the velocity of the stochastic fronts is shown to depend on the channel width in a simple and interesting way, in contrast to fronts in the deterministic MFE.  Thus, fluctuations alter the behavior of these fronts in an essential way.  These affects are shown to be partially captured by introducing a density cutoff in the reaction rate.  Some of the predictions of the cutoff mean-field approach are shown to be in quantitative accord with the stochastic results. 
\end{abstract}

\maketitle

\section{\label{sec:level1}Introduction}

Several well known processes in spatially extended systems exhibit fronts that propagate through space.  Most of these processes that have been considered to date occur in media in which the governing dynamics are spatially uniform.  Recently, however, some interesting findings have been made concerning fronts propagating in systems with spatially heterogeneous dynamics.  In particular, the simple infection model   $A+B \rightarrow 2A$ on a lattice with equal hopping rates and a linear reaction rate gradient has been studied.  Two versions of this system have been examined in some detail \cite{Kessler:nz,Cohen:2005dn} : one in which the gradient is defined with respect to the medium itself (the "absolute gradient"), and another in which the gradient is defined relative the the front's interface and travels along with the front (the "quasi-static gradient").  One can imagine numerous systems that can be described by the absolute gradient, e.g. a chemical reaction occurring in a temperature gradient.  The quasi-static gradient is more analytically tractable and also arises naturally in models of biological evolution\cite{Rouzine:2003no,Tsimring:1996bu}.  

The usual way to analytically study a system with a propagating front, such as the infection model mentioned above, is within a mean field, reaction-diffusion framework.  The simplest MF analog to our infection model is the usual Fisher equation \cite{Fisher:1937mv}, with a spatially varying reaction rate:

\begin{equation}
\partial\phi/\partial t = D\nabla^2\phi + r(x)\phi(1-\phi) 
\label{gen-rd}
\end{equation}

For our simple infection model, Eq.\ref{gen-rd} (the "naive MFE") fails to to capture many of the qualitative aspects of the stochastic problem with either absolute of quasi-static gradients.  These failures, as well as many other many other issues involving the MF description of similar front propagation problems, are largely remedied by introducing a cutoff factor in the reaction term:

\begin{equation}
\partial\phi/\partial t = D\nabla^2\phi + r(x)\phi(1-\phi) \theta(\phi - \phi_{c})
\label{eqofmo}
\end{equation}

This term causes the reaction rate to abruptly drop to zero in regions far into the front's leading edge where $\phi$ drops below a critical level $\phi_{c}$, and is meant to mimic the leading order effect of finite number fluctuations in the stochastic process.  In other words, the discrete nature of individual particles implies that a sufficiently small density field $\phi<\phi_c \sim 1/N$ corresponds, in an average sense, to zero particles present and thus zero reaction rate. 

In this study, we turn our attention to the quasi-static gradient in two dimensions and ask how finite number fluctuations and the related cutoff approach affect the stability of planar fronts propagating in a rectangular channel.  In what follows we will see that in contrast to the predictions of the naive MFE, the results of stochastic simulations point to unstable planar fronts.  Furthermore, once again the cutoff term will rescue the effectiveness of the mean field description of the stochastic process.  We first study the cutoff mean-field 
equations, both numerically
and analytically, showing the instability.  We then turn
to the stochastic model, demonstrating the instability there as well. An appendix contains details about the numerics.

\section{Mean Field Stability Calculation}
The full equation of motion governing the quasi-static gradient in the MF cutoff framework is Eq. \ref{eqofmo} with

$$
\begin{array}{c}
\phi \equiv N_{A}/N
\\
r(x) \equiv \textrm{max}(r_{min}, r_{o}+\alpha (x-\bar{x}))
\\
\bar{x} \equiv \frac{1}{b}\int \phi (x,y) dxdy
\\
\phi_{c} = k/N
\end {array}$$

Here $x$ is the direction parallel to the channel's long axis, $N$ is the equilibrium number of $A$ particles per lattice site, and $k$ is some $O(1)$ fitting parameter.  $\bar{x}$ serves to define the interface position of the front by essentially comparing the front's profile to a step function, and $b$ is the cross-channel width.  $r_{min}$ merely serves to keep the front stable and plays no important role, as its effects are only felt far from the leading edge.  In the  context of biological evolution,  $\phi$ represents the fraction of individuals in a population with a given fitness $x$.  If the size of the population is fixed, the growth rate of individuals with a particular fitness is proportional to $x - \bar{x}$.  The diffusion term represents "genetic drift", and the dynamics of the system corresponds to the population evolving towards greater mean fitness.  

\begin{figure}
\includegraphics[width=6in]{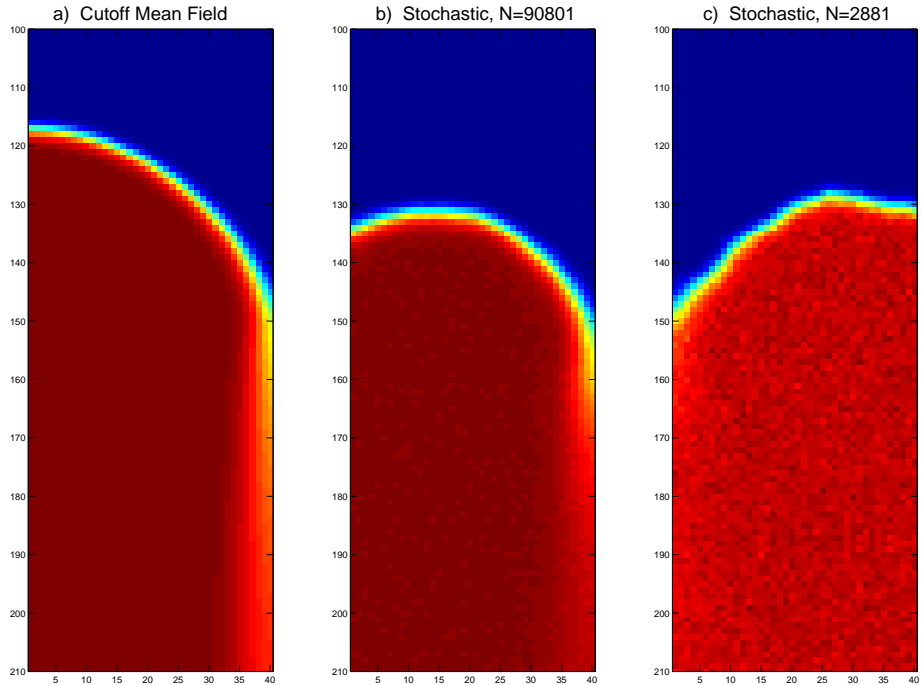}
\caption{Snapshots of the growing finger for the cutoff MFE compared to that
for the stochastic model. The parameters are $D=1$, $r_0=6$, $\alpha=0.3$.
a)The cutoff MFE with $k/N=8.7\cdot 10^{-5}$, b) the stochastic model with $N=90801$, 
c) the stochastic model with $N=2881$.}
\label{fingerfig}
\end{figure}

\begin{figure}
\includegraphics[width=6in]{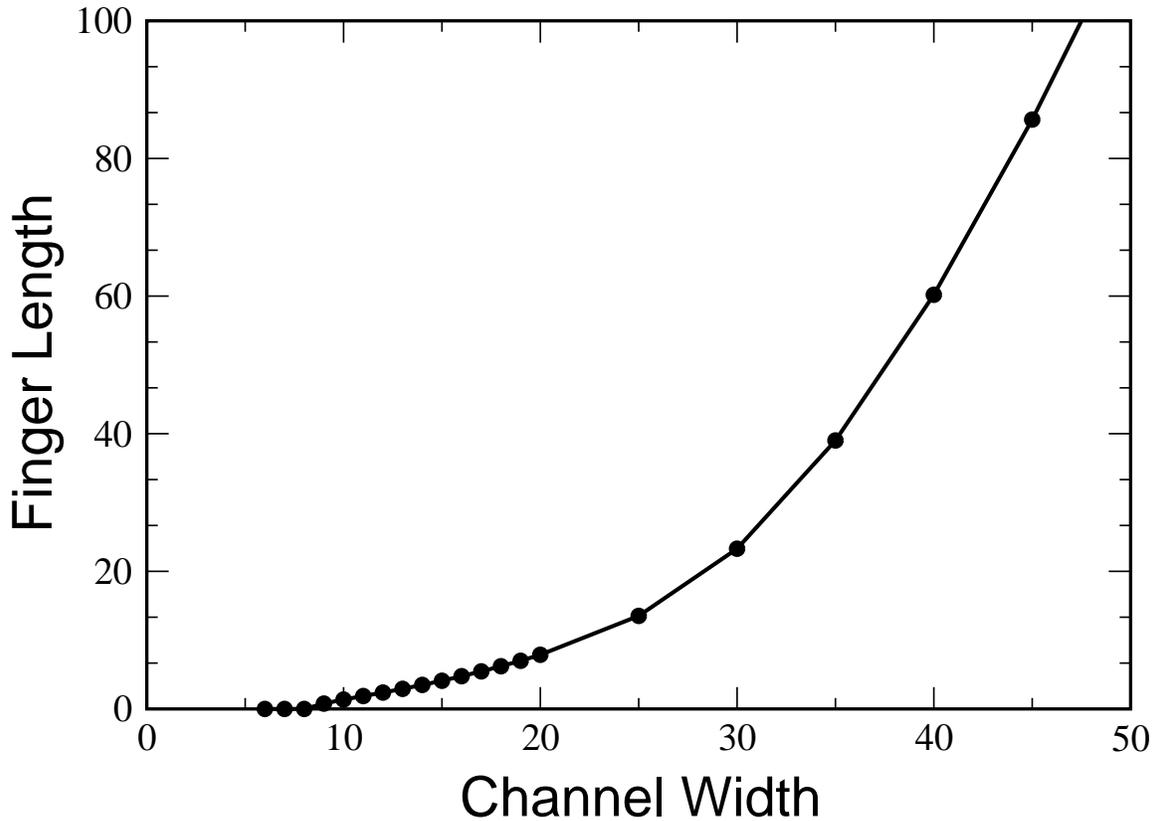}
\caption{Finger length (as defined in the text) versus channel width for the cutoff MFE. The parameters
are: $D=1$, $r_0=6$, $\alpha=0.3$, $k/N=8.7\cdot 10^{-5}$.}
\label{widthfig}
\end{figure}

Direct numerical integration of the time dependent Eq. \ref{eqofmo} shows that planar fronts are in fact unstable.  For a sufficiently wide channel, perturbed planar fronts develop into long, though finite,  fingers whose length increases with increasing channel width. An example of such a finger is shown in Fig. \ref{fingerfig}.  We see that there is a deep narrow ``notch'' on the trailing side of
the finger, so that the width of the interface is much greater here than for the
rest of the finger. Defining the finger length by $\int \left[\phi (x,0) -\phi(x,b)\right]dx$, the data for finger 
length versus channel width is presented in
Fig. \ref{widthfig}. We now turn toward an analytic understanding of this result.  

Due to the translational invariance of the system, it is natural to investigate first steady-state propagating
planar front solutions.
Plugging into Eq. (\ref{eqofmo}) the traveling wave form, $\phi_{0}(x,y,t) = \phi_{0}(x-vt)$, with velocity $v$, we obtain
\begin{equation}
D\phi_{0}''+ v\phi_{0}' + r(z)\phi_{0} (1 - \phi_{0})\theta\left(\phi_{0}-\frac{k}{N}\right) = 0
\label{ss}
\end{equation}
in terms of the comoving coordinate $z \equiv x-vt$.  A quick analysis of the linearized version of Eq. (\ref{ss}) provides insight into the role of the cutoff.  As $z\rightarrow -\infty$, $\phi_{0} \rightarrow 1$.  Linearizing  around $\phi_{0}=1$, we find two exponential solutions, but one must be discarded since it decreases with increasing z.  Similarly, as $z \rightarrow \infty$, $\phi_{0} \rightarrow 0$.  Linearizing around $\phi_{0}$ = 0, in the region past the cutoff, once again we find only one acceptable, decaying solution.  This leaves our solution with a total of two undetermined constants.  Fixing translational invariance reduces this number to one.  Requiring continuity of $\phi_{0}$ at the cutoff determines the remaining coefficient, and continuity of $\phi_0'$ determines the velocity.  Thus, mathematically, the cutoff fixes the velocity by overdetermining the boundary conditions, i.e. converting Eq. (\ref{ss}) into an eigenvalue problem.  An analysis for large $N$ yields the leading order result~\cite{Kessler:nz}:
\begin{equation}
v=\left[24D^2 \alpha\ln{(N/k)}\right]^{1/3}
\end{equation}
In the limit $k/N \rightarrow 0$ we regain the naive MF approach, in which $v\rightarrow \infty$.  Thus the naive MF and the cutoff MF predict qualitatively different results with respect to velocity.  Not surprisingly, stochastic fronts in fact approach a (finite) steady-state velocity, in line with the cutoff MF.  

Turning now to 2D fronts, we wish to study the linear stability of the planar solution to transverse perturbations.
We write $\phi(x,y,t)=\phi_0(z) + \tilde{\phi}(z,y,t)$ and linearize Eq. (\ref{eqofmo}) with respect to $\tilde{\phi}$. The
invariance of the system with respect to translations in time and the transverse spatial direction $y$ implies $\tilde{\phi}(z,y,t) =  e^{\omega t} e^{iqy}\eta(z) $.  The governing equation for $\eta(z)$ is then:
\begin{equation}
D\eta'' + v\eta' + \eta r(z)[(1-2\phi_{o})\theta(\phi_{0}-k/N) + \phi_{0}(1-\phi_{0})\delta(\phi_{0}-k/N)] = \Omega\eta
\label{stab}
\end{equation}
with
\begin{equation}
\Omega \equiv Dq^2+\omega
\label{omega}
\end{equation}
The delta function arises from differentiating the step function and is due to the shift in $z_{cut}$ caused by the perturbation $\tilde{\phi}$. We have assumed here that $q\ne 0$ so that $\int \tilde{\phi}(z,y,t)dy=0$.  The 
case $q=0$ has to be treated separately, but in any case the least stable mode should be the translation mode
with $\Omega=0$. Notice that Eq. (\ref{omega}) implies a simple stabilizing quadratic dependence of the growth rate $\omega$ on $q$. Thus the
least stable mode is that with the smallest non-zero $q$, which, assuming a zero-flux sidewall boundary condition, is
$q_{min}=\pi/b$.  This implies a minimum channel width $b^{*}$ below which even the longest wavelength mode has too much curvature for any instability to exist:
\begin{equation}
b^* = \pi \sqrt{\frac{D}{\Omega_{max}}}
\label{bstar}
\end{equation}
where $\Omega_{max}$ is the largest eigenvalue of the stability operator, Eq. (\ref{stab}).

Like the steady-state problem,  insight can be gained into Eq. (\ref{stab}) by considering the boundary conditions at $z \rightarrow \pm \infty$.  We require that $\eta \rightarrow 0$ as $z \rightarrow -\infty$.  As $\phi_0 \sim 1$ in this region, we find two exponential solutions for $\eta$: one growing with increasing z and the other decaying.  The latter must of course be excluded.  If we perform the same procedure past the cutoff, as $z\rightarrow \infty$, we find two decaying modes.  However, one of the modes decays more slowly than the steady-state solution and  thus dominates it for sufficiently large $z$.  This is unacceptable behavior for a perturbation and therefore this solution is discarded.  Thus, our solution has two arbitrary constants, one of which may be chosen arbitrarily since Eq. (\ref{stab}) is linear in $\eta$.  The remaining constant is fixed by requiring continuity of $\eta$ at the cutoff.  Matching $\eta'$ at the cutoff determines the eigenvalues $\Omega$.  Thus, once again the cutoff has played a central role in determining the problem's interesting quantities.

As with the steady-state problem, we can make analytic progress in  the limit of large $N$. In this case, the cutoff is at large $z$, in the region where $\phi_0$ is small.
If we consider Eq. (\ref{stab}) in the region where $\phi_0 \ll 1$ and $z<z_{cut}$, and fix the translation invariance by setting $\bar{z}=0$ for the unperturbed state, we obtain

\begin{equation}
D\eta'' + v\eta' + \eta (r_0+\alpha z) = \Omega \eta
\label{eta-ODE-appx}
\end{equation}

Up to a similarity transformation, this is the Airy equation, with the general solution
\begin{equation}
\eta=e^{-\frac{vz}{2D}}\left[A \textrm{Ai}\left(\frac{\Gamma-z}{\delta}\right) + B \textrm{Bi}\left(\frac{\Gamma-z}{\delta}\right)\right]
\label{airybairy}
\end{equation}

with
\begin{center}
$\Gamma \equiv \frac{v^2/4D - r_0 + \Omega}{\alpha}$
\\
$\delta \equiv \left(\frac{D}{\alpha}\right)^{1/3}$
\end{center}

We argue that the Bi term must vanish by considering the large $v$ limit of Eq. (\ref{stab}) and matching onto Eq. (\ref{airybairy}).  As shown in \cite{Cohen:2005dn}, in the large $v$ limit, the diffusion term in Eq. (\ref{ss}) can be ignored, and the solution in the region where $\phi_0\ll1$ is

\begin{equation}
\eta \sim e^{-\frac{1}{v}[(r_0-\Omega)z + \frac{1}{2}\alpha z^2]}
\label{bulk-ss}
\end{equation}

Expansions of $Ai$ and $Bi$ for large argument show that the diffusionless result Eq. (\ref{bulk-ss}) matches onto 
Eq. (\ref{airybairy}) only if the $Bi$ term is absent.  The constant $A$ may be arbitrarily set to unity since the problem is linear.  
Thus we have for $ z\lesssim z_{cut}$
\begin{center}
$\eta=e^{-\frac{vz}{2D}}\textrm{Ai}\left(\frac{\Gamma-z}{\delta}\right) $       
\end{center}
We have to match this result to the simple exponential solution for $z>z_{cut}$.  Thus,
\begin{equation}
e^{-\frac{vz_{cut}}{2D}}\textrm{Ai}\left(\frac{\Gamma-z_{cut}}{\delta}\right) =C e^{\left[-\frac{vz_{cut}}{2D}\left(1+\sqrt{1+\frac{4\Omega_0D}{v^2}}  \right)\right]}
\end{equation}
The derivative of $\eta$ must also match properly at the cutoff.  Looking back to Eq. (\ref{stab}), we see that the delta function term causes a discontinuity in $\eta'$ at $z_{cut}$:
\begin{equation}
\eta'_{right}-\eta'_{left} = \frac{r(z_{cut})}{D}\frac{\frac{k}{N}(1-k/N)}{\phi_0'(z_{cut})}= -\frac{r(z_{cut})}{v}(1-k/N)
\end{equation}
Computing the derivatives in (12) and dividing by (11), we obtain
\begin{equation}
\frac{v}{2D}\sqrt{1+4D\Omega/v^2} - \frac{1}{v}(r_{0}+\alpha z_{cut})(1-k/N)
=\frac{1}{\delta}  \frac{Ai'(\frac{\Gamma(\Omega) - z_{cut}} {\delta})} {Ai(\frac{\Gamma(\Omega) - z_{cut}} {\delta})}
\label{match}
\end{equation}
This equation determines $\Omega$ if the quantities $v$ and $z_{cut}$ are known.  Now, for large $v$ the LHS of (12) is also large.  For the RHS to balance it, the Airy function in the denominator must be small.  Thus, $\frac{\Gamma-z_{cut}}{\delta} \approx \xi_0$, where $\xi_0\approx-2.3381$ is the first zero of the Airy function.  For the value of the cutoff, we quote another result from \cite{Kessler:nz} obtained by matching the linearized steady-state equation at the cutoff:
\begin{equation}
z_{cut} \approx\frac{v^2/4D - r_0}{\alpha}-\xi_0 \delta -2D/v
\end{equation}
Plugging this expression into Eq. (\ref{match}), expanding around $\xi_0$, and dropping higher order terms, we obtain the leading order result valid for large $v$:

 \begin{equation}
 \Omega=\frac{2D\alpha}{v}
\label{final-omega}
  \end{equation}

\begin{figure}[ht]
\begin{center}
\includegraphics[width=6in]{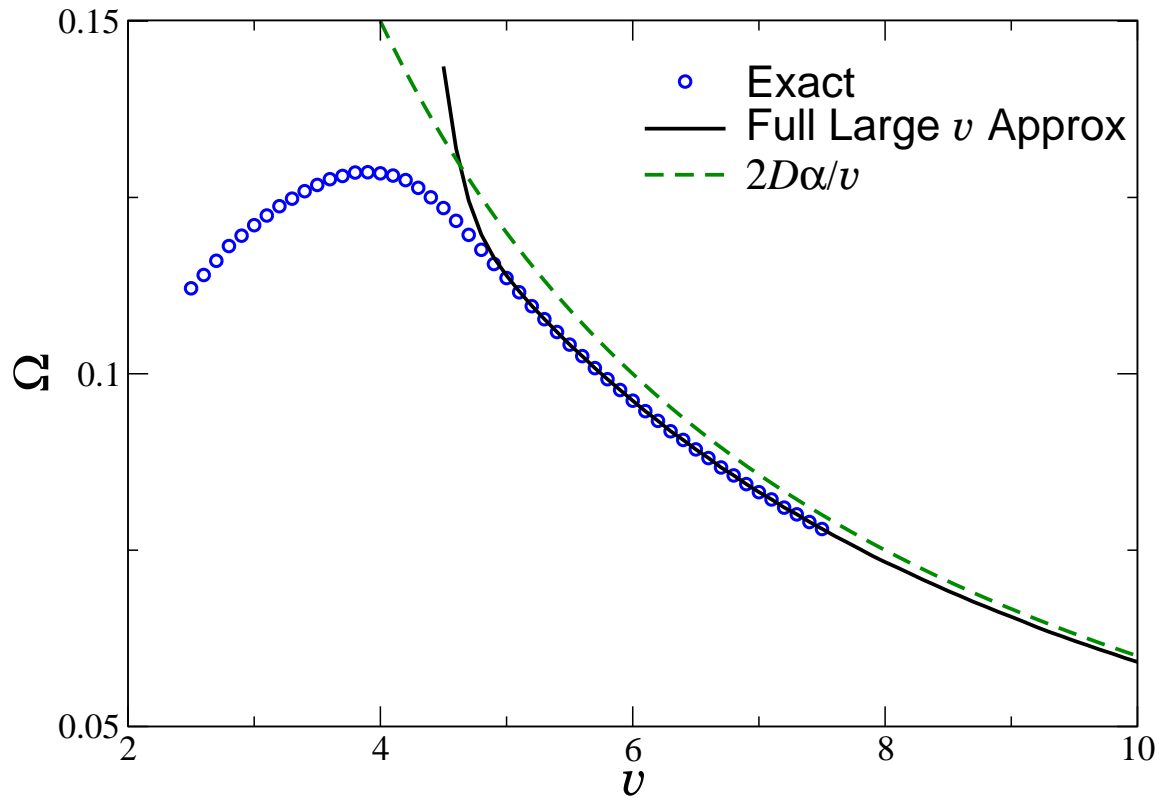}
\caption{(Color online)The circles represent the exact numeric solution of Eq. (\ref{stab}). The solid line is the exact numeric solution of Eq. (\ref{match}) which is itself a large $v$ approximation.  Exact numerically generated values of $N$ and $z_{cut}$ were used to generate this approximation.  The dashed curve is the analytic approximation Eq. (\ref{final-omega}). The parameters are $D=r_0=1$, $\alpha=0.3$.}
\label{fig1}
\end{center}
\end{figure}
 
Here we see the interesting result that plane fronts become stable in the limit as $v\rightarrow \infty$, i.e. $N \rightarrow \infty$, i.e. the cutoff disappears.  Eq. (\ref{final-omega}) can be interpreted as saying the that $\Omega$ is proportional to the ratio of the diffusive length scale (D/v) to the length scale over which the rate changes appreciably ($1/\alpha$).  Thus, heuristically incorporating the effects of finite number fluctuations qualitatively changed the system's stability properties by limiting the front's velocity, which in turn makes the diffusive length scale finite.  This fluctuation induced instability is similar to that in \cite{Kesslar:1998pu}, where it was found that that a coupled reaction diffusion system with no reaction gradient, but with unequal diffusion coefficients, is unstable with a cutoff but stable without one.  Furthermore, Eq. (\ref{final-omega}) shows that the fronts become stable as $\alpha \rightarrow 0$, for any value of $N$.  This is analogous to the stability shown in \cite{Kesslar:1998pu} in the case of equal diffusion coefficients.  



Thus, once again, the presence of the cutoff qualitatively changes the simple mean field predictions.  If the cutoff approach is to mimic the effects of finite number particle fluctuations, we should expect to see some analog of this front instability in the stochastic, discrete infection model discussed earlier, to which we now return.

\section{Stochastic System}
We ran simulations in which the lower rectangular portion of the channel was initially populated with $N$ type $A$ particles per lattice site and the upper rectangular portion of the channel which was populated by $N$ type $B$ particles per lattice site.  During each time step, a binomially distributed random number of particles hops to adjacent sites.  Furthermore, $A$ particles probabilistically cause some $B$ particles to change into $A$ particles .  The reaction probability and hopping rates were chosen so that the discretized, stochastic equation for $\Delta N_A$ reduces to Eq.\ref{gen-rd}  (with the quasi-static form for $r(x)$) when the expectation value is taken in the small time, small lattice spacing limit.   In particular, for the hopping probability we took $P_{hop}\sim D\frac{ dt}{l^2}$, where $dt$ is the simulation timestep and $l$ is the lattice spacing.  The number of particles reacting during each timestep was chosen as a binomially distributed random variable with mean $1-(1-\frac{r(x)dt}{N})^{N_B}$ drawn $N_A$ times.  

Results of simulations depended crucially on the width of the channel $b$ and, to a lesser extent, the equilibrium number of particles per lattice site $N$.  Initially uniform planar fronts were roughened by random fluctuations.  These fluctuations, especially those occurring near the interface, often developed into a more organized, longer wavelength deviations from planarity.  Naturally, these fluctuations occurred more frequently in runs with smaller $N$.  For $b\lesssim b^*$, diffusion tended to flatten these deviations back to planarity over a characteristic time scale $t_{dev}$.  Although  $t_{dev}$ is difficult to measure quantitatively, it is qualitatively clear that it increased with increasing $b$.  Not surprisingly, as $t_{dev}$ and $b$ increased, so did the amplitude of the associated deviations.  The divergence of $t_{dev}$ corresponds to the aforementioned phase transition predicted by the cutoff MF approach in Eq. \ref{bstar}.  For $b\gg b^*$ $t_{dev}$ did in fact appear to diverge, resulting in patterns that
resemble fingers, rather that random fluctuations.  As $b$ is increased further,the finger-like nature of the pattern becomes increasingly pronounced.  This
trend is of course clearly the larger $N$ is.  In Fig. \ref{fingerfig}, we present the stochastic finger for two different values of $N$, alongside the
pattern produced by the cutoff MF.  The larger $N$ corresponds to the
value of the cutoff chosen for the MF simulation. The overall similarity between
these two figures is clear.  What is also striking is the relatively large
effect of noise, even though $N$ is quite large, pointing out once more the
crucial importance of fluctuations in the system.  The fluctuations are of
course even more pronounced in the smaller $N$ system.

The random roughening of the interface makes the identification of the point
of onset of the instability in the stochastic system somewhat subtle, given
the small size of the pattern near onset.  In order to compare the stochastic system to the MF prediction, we need a way to distinguish this random roughening of the interface from the genuine pattern forming mechanism discussed in the previous section.  We devised two tests whose results we believe demonstrate the existence of an instability in the stochastic system.  

\begin{figure}[ht]
\begin{center}
\includegraphics[width=5in]{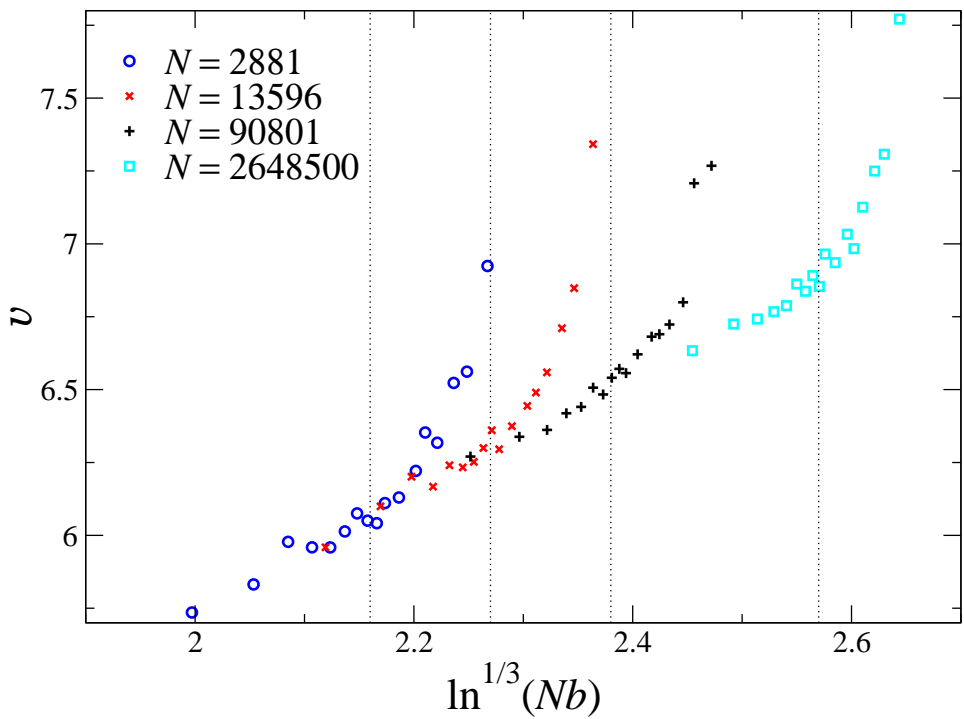}
\caption{(Color online)Evidence of the transition to instability.  The linear envelope of the different curves demonstrates the $N$ renormalization $N_{eff}\sim Nb$ required by widening the channel.  The dotted lines show the $b^{*} $resulting from the corresponding deterministic, MF cutoff simulation ($k$=.25).  Each data point represents an ensemble average of three long trials (100 time units each). The parameters
are $D=1$, $r_0=6$, $\alpha=0.3.$}
\label{fig2}
\end{center}
\end{figure}

Both of these tests exploit the predicted transition between stable and unstable states that occurs when the channel width exceeds a critical value $b^*$, as stated in (6).  First, we measured the ensemble averaged  velocity of the mean interface $\bar{z} \equiv \sum_{i,j} \frac{A(i,j)}{Nb}$ as a function of $b$ (Fig. \ref{fig2}).  The increasing trend along the envelope of the different curves can be understood as a result of wider interfaces presenting an effectively larger number of particles $N_{eff}$.  In fact, since in the steady-state, $v\sim (\ln N)^{1/3}$, we see from the figure the remarkably simple result $N_{eff}\sim Nb$.  This simple dependence continues until $b$ approaches $b^*$ (dotted vertical lines), where the velocity suddenly increases.  This increase can be understood as a result of the system spending much of its time in a configuration in which one side of the interface significantly leads in front of the other.  The lagging side then effectively stalls while the leading side is in a region of large reaction rate, and thus propagates quickly.  The overall effect is an increase in the velocity averaged over the width of the channel.  The fact that the change occurs so near the $b^*$ calculated earlier suggests that the cutoff approach is effectively capturing the stochastic dynamics.

As a second test, we plotted the mean roughness of the interface $W$ vs $b$ (figure 3).  $W$ is defined in the standard way
\begin{center}
$W^2\equiv \langle \overline{\left[\left(\sum_{j}A(i,j)/N\right) - \bar{z} \right]^2 }\rangle$
\end{center}
where $\langle \rangle$ denotes ensemble average and the bar denotes average over the transverse direction.  For $b<b^*$ we see power law scaling reminiscent of that discovered by Kardar, Parisi, and Zhang \cite{Kardar:1986ty} for a growing interface.  However, the data shows no sign of a universal exponent.  It may be that the very weak
stability of the interface near $b^*$ is responsible for a long crossover.  This issue clearly requires more
extensive study.  For a fixed $b$, $W$ decreases with increasing $N$,  consistent with the hypothesis that interface roughness is noise driven in this regime.   However, for $b\gtrsim b^*$ this simple dependence is lost.  The curves converge near $b^*$, showing that particle number and its associated noise are no longer the relevant factor in determining interface roughness.  Past this intersection, there is no apparent correlation between W and N.  We interpret this as a crossover from noise driven interface roughness to gradient driven pattern formation occurring very near the $b^*$ predicted from the cutoff MF approach. 

\begin{figure}[ht]
\begin{center}
\includegraphics[width=5in]{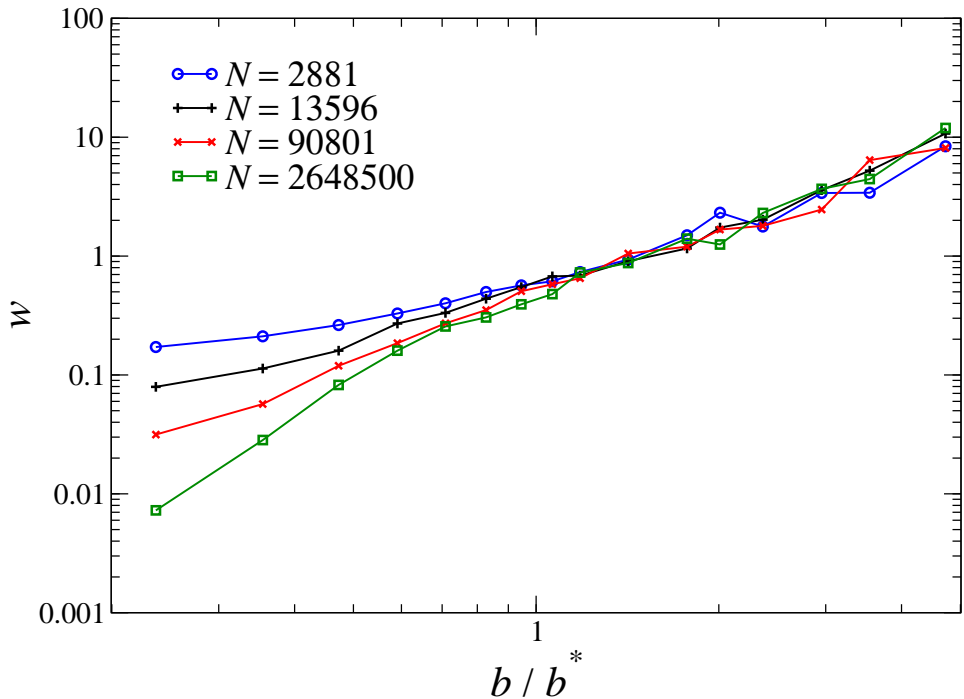}
\caption{(Color online)Noise driven roughness scaling for $b<b^*$, gradient driven scaling thereafter.  Each data point represents an ensemble average of three long trials (100 time units each). Parameters are $D=1$, $r_0=6$, $\alpha=0.3$.}
\label{fig3}
\end{center}
\end{figure} 

Thus, the cutoff MF approach is quantitatively successful in predicting the transition of the width and velocity observed in ensemble averaged stochastic fronts when the channel is widened.  In contrast, the naive MF approach predicts no such transition and an infinite steady-state velocity, in qualitative disagreement with the simulation results.  The cutoff MF approach also predicts the velocity of the average interface for $b<b^*$, provided we take $N \rightarrow Nb$.  

Another aspect of the stochastic system which one would like to predict is the
ensemble-averaged shape. We find that qualitatively this behaves as expected;
namely, for small channel width the average shape is flat, and above the
critical width, a nontrivial shape is apparent.  The amplitude of the averaged
pattern continues to increase with increasing width.  However, 
we do not know how to quantitatively relate the average pattern to the
results of the cutoff MF equations.  One obvious impediment is the fact that
the stochastic system switches parity at random, with the right and left sides
alternating as the leading edge.  Thus, a naive ensemble average produces a
shape which is highest in the center, clearly at odds with the deterministic
calculation.  Another aspect of the problem that we would like to correlate
with the deterministic calculation is the growth rate of the pattern near
onset.  This problem is also difficult because the width, measured in the usual way, consists of a contribution from noise driven roughening and one from pattern formation.  Clearly, the usual MF approach can only make predictions about the contribution from patterning and thus the stochastic results and MF predictions are intrinsically difficult to compare.  Even though the noise decreases with
increasing $N$, the dominance of the dynamics by the leading edge where
fluctuations are unavoidably present makes this a nontrivial task, even at
large $N$.  These questions remain challenges for the future.  


\begin{acknowledgments}

The work of HL and CSW was supported in part by the NSF PFC-sponsored Center for Theoretical Biological Physics (Grants No. PHY-0216576 and PHY-0225630).
The work of DAK was supported in part by the Israel Science Foundation.

\end{acknowledgments}
\appendix
\section{Details Concerning Numerical Integration}
In order to determine the spectrum $\Omega_0(v)$ we numerically integrated Eq. 5, and this required numerically integrating Eq. 3.  Integration of Eq. 3 was initialized in the bulk state, from the left, where we defined $z \equiv 0$.  We took set $v$ arbitrarily, took $\phi_{0}(0)=.99$, and calculated $\phi_{0}'(0)$ from the solution to the version of Eq. 3 linearized around $\phi_0 \approx 1$
\begin{center}
$ \phi_0'(0) = -\frac{.01v}{2D} \left(-1+\sqrt{1+\frac{4Dr_{init}}{v^2}}\right)$
\end{center}
$r_{init}$, which we set to one, differs from the previously defined $r_0$ in that it fixed the rate at a fixed location in the bulk state rather than where $\phi_0=1/2$.  Integration terminated in the neighborhood of the cutoff, half way between timesteps where  $\phi_0'/\phi_0$ crosses $-v/D$. $N$ was then read off from the relation $N = -\frac{vk}{D\phi_0(z_{cut})}$ and the value of the cutoff was recorded for subsequent numerics.  This value is measured relative to the bulk position where $\phi_0=.99$, not relative to $\bar{z}$, where $\phi_0 = 1/2$.  All of this was done using ode45 from MATLAB, with a maximum stepsize of .001.  

The solution for $\phi_0$ appears as a coefficient in Eq. (\ref{stab}), and was incorporated into the ODE integration scheme with a cubic spline.  Numerical integration of Eq. (\ref{stab}) was initialized at $z=0$ with some trial $\Omega_0$, $\eta(0)=1$ and 
\begin{center}
$\eta'(0) =\frac{v}{2D} \left( -1+\sqrt{1+\frac{4D(r_{init}+\Omega_0)} {v^2}}\right)$
\end{center}
which follows from Eq. (\ref{stab}) if we plug in $\phi_0 \approx 1$.  Integration terminated at the $z_{cut}$, where we checked if Eq. (\ref{match}) was satisfied with the trial $\Omega$.  This procedure was iterated with a root solver while varying  $\Omega$ until Eq. (\ref{match}) was satisfied.  Each integration of Eq. (\ref{stab}) was done over 1000 timesteps with a fourth order Runge-Kutta ODE solver with fixed step size, meant to facilitate incorporation of the spline.  

This yielded the exact numeric solution presented in Fig. 1.  Our "exact analytic approximation" presented in Fig. 1 is just the solution to Eq. \ref{match} obtained with a root solver.  The required values for $N$ and $z_{cut}$ were obtained from the previous integration of Eq. (\ref{ss}), and we took $r_0=1$.  Since we dropped  the term involving $\phi_0$ in Eq. \ref{eta-ODE-appx}, this equation is insensitive to the precise definition of $\bar{z}$, e.g. it could be defined naturally as the coordinate where $\phi_0 = 1/2$ or it could be defined out of numerical convenience as the coordinate where $\phi_0 = .99$.  This insensitivity explains why the results agree so well despite the fact that $r_0$ and $r_{init}$ are not defined in the same way.


\bibliography{frontrefs}

\end{document}